\documentclass[pra,showpacs,twocolumn,aps,superscriptaddress,a4paper]{revtex4-1}
\usepackage{dcolumn,amssymb,amsmath,amsfonts,graphicx,latexsym,color}
\usepackage{epstopdf}

\begin{document}


\title{Polaron in a $p+ip$ Fermi topological superfluid}

\author{Fang Qin}
\email{qinfang@ustc.edu.cn}
\affiliation{CAS Key Laboratory of Quantum Information, University of Science and Technology of China, Chinese Academy of Sciences, Hefei, Anhui 230026, China}
\affiliation{CAS Center For Excellence in Quantum Information and Quantum Physics, University of Science and Technology of China, Hefei, Anhui 230026, China}
\author{Xiaoling Cui}
\email{xlcui@iphy.ac.cn}
\affiliation{Beijing National Laboratory for Condensed Matter Physics, Institute of Physics, Chinese Academy of Sciences, Beijing 100190, China}
\affiliation{Songshan Lake Materials Laboratory,  Dongguan, Guangdong 523808, China}
\author{Wei Yi}
\email{wyiz@ustc.edu.cn}
\affiliation{CAS Key Laboratory of Quantum Information, University of Science and Technology of China, Chinese Academy of Sciences, Hefei, Anhui 230026, China}
\affiliation{CAS Center For Excellence in Quantum Information and Quantum Physics, University of Science and Technology of China, Hefei, Anhui 230026, China}

\date{\today}

\begin{abstract}
We study polaron excitations induced by an impurity interacting with a two-dimensional $p+ip$ Fermi superfluid. As the Fermi-Fermi pairing interaction is tuned, the background Fermi superfluid undergoes a topological phase transition. We show that such a transition is accompanied by a discontinuity in the second derivative of the polaron energy, regardless of the impurity-fermion interaction. We also identify a polaron to trimer crossover when the Fermi superfluid is in the strongly interacting, thus topologically trivial, regime. However, the trimer state is metastable against the molecular state where the impurity binds a Bogoliubov quasiparticle from the Fermi superfluid. By comparing the polaron to molecule transition in our system with that of an impurity in a noninteracting Fermi sea, we find that pairing interactions in the background Fermi superfluid effectively facilitate the impurity-fermion binding. Our results suggest the possibility of using the impurity as a probe for detecting topological phase transitions in the background; they also reveal interesting competitions between various many-body states in the system.
\end{abstract}

\maketitle

\section{Introduction}\label{1}

The study of topological matter has attracted much attention in recent years~\cite{Kane2010,Qi2011,Shen2012,Bernevig2013,Ryu2016,Wen2017,Cooper2018,Zhu2018}. Whereas topological bands and phases in noninteracting systems are well understood by now, the interplay of interaction and topological bands is still under extensive study~\cite{Qi2008,Galitski2010,Gurarie2011,Galitski2011,Wen2014,Senthil2014,You2014,You2015,Zheng2015,Xu2016,Vanhala2016,Pan2016,Jian2018,Zhang2018,Rachel2018}.
A bottom-up approach here concerns an impurity interacting with a topologically nontrivial background. As the impurity is dressed by quasiparticle excitations of its environment, it can also acquire some topological features through interaction. In previous studies, both immobile and mobile impurities have been considered in topological environments~\cite{TopologicalHu2013,TopologicalPolaron2016,TopologicalBruun2018}, with the latter giving rise to polarons where the impurity moves around dragging particle-hole excitations along. Such a scenario is especially relevant to ultracold atomic gases, where key elements such as topological bands~\cite{ExpTbands2014,ExpTbands2015,ExpTbands2016,ExpTbands2018}, tunable interactions~\cite{ExpFeshbach2010}, and polaron excitations~\cite{ExpFermiPolaron2009,ExpFermiPolaron2D2012,ExpFermiPolaron3D2012,ExpFermiPolaron3D2017,ExpBosePolaron3D20161,ExpBosePolaron3D20162,ExpBosePolaron3D20163,ExpBosePolaron3D2018,Mistakidis20181,Mistakidis20182} are all experimentally accessible. While polarons serve as a bridge connecting few- and many-body physics~\cite{Chevy2006,Chevy2007,Combescot2008,Chevy2010,Bruun2014,SOCpolaron2014,Bruun2D2011,Parish2D2011-1,Parish2D2013,SOC2D2012,SOCpolaron2016,Recati2D2011,Parish2D2011-2,Schmidt2D2012,Yi2015,Nishida2015}, they provide a unique angle for understanding topological systems in the presence of interaction.

In this work, we study an impurity immersed in a two-dimensional $p+ip$ Fermi superfluid. As the $p$-wave Fermi-Fermi interaction is tuned, the superfluid becomes topologically nontrivial in the weak-coupling regime with $\mu>0$, where $\mu$ is the chemical potential of fermions. In contrast, in the strong-coupling regime with $\mu<0$, the superfluid is topologically trivial. A topological phase transition occurs at $\mu=0$, where the system becomes gapless~\cite{Machida2008,Alicea2012}. Assuming a tunable impurity-fermion interaction, we calculate the energy, the impurity residue, and the wave function of the polaron state, where the impurity is dressed by a pair of Bogoliubov quasiparticles. We find that as the background Fermi superfluid is tuned across the topological phase transition, a discontinuity emerges in the second derivative of the polaron energy, which is consistent with the order of the topological phase transition. On the other hand, when the impurity-fermion interaction is tuned, the polaron state can cross over into a trimer state, as the impurity residue decreases rapidly within a finite range of interaction strength. Interestingly, such a polaron to trimer crossover only occurs when the background Fermi superfluid is in the topologically trivial strong-coupling regime. Further, as the impurity-fermion interaction increases, the polaron state can become metastable against a molecular state, where the impurity forms a local bound state with a single Bogoliubov quasiparticle. In the strong-coupling regime, the polaron to molecular transition occurs before the polaron-trimer crossover, such that the trimer state is metastable. In the weak-coupling limit, the polaron state remains the ground state.

The paper is organized as the following: In Sec.~\ref{2}, we describe the system configuration and give the model Hamiltonian. In Sec.~\ref{3}, we characterize the polaron state using a variational approach. We characterize the molecular state as well as the polaron to molecule transition in Sec.~\ref{4}. Finally, we summarize in Sec.~\ref{5}.

\section{Model}\label{2}

We consider an impurity interacting with a two-dimensional $p+ip$ Fermi superfluid. The Hamiltonian is
\begin{align}
H-\mu N
&= \sum_{{\bf k}} (\epsilon_{\bf k} - \mu) a^{\dag}_{{\bf k}} a_{{\bf k}} + \sum_{{\bf k}}\epsilon_{{\bf k}}^{b}b_{{\bf k}}^{\dagger} b_{{\bf k}} \nonumber\\
&~~ + \frac{1}{2}\sum_{{\bf k}}\left( \Delta_{\bf k}a_{{\bf k}}^{\dagger}a_{-{\bf k}}^{\dagger} + \Delta_{\bf k}^{*}a_{-{\bf k}}a_{{\bf k}} \right) \nonumber\\
&~~ + \frac{g_{\rm{fi}}}{V}\sum_{{\bf k},{\bf k}',{\bf q}}a_{{\bf k}}^{\dagger}b_{{\bf q}-{\bf k}}^{\dagger}b_{{\bf q}-{\bf k}'}a_{{\bf k}'},
\end{align}
where $N$ is the total number of spinless fermions in background superfluid, $\mu$ is the chemical potential of the fermions in the background, $a_{{\bf k}}$  and $b_{{\bf k}}$ are,  respectively, the annihilation operators for the background $p$-wave superfluid fermions and the impurity atom, $\epsilon_{{\bf k}} = k^2/(2m)$ is the kinetic energy of the spinless fermions with mass $m$, $\epsilon_{{\bf k}}^{b} = k^2/(2m_b)$ is the kinetic energy of the impurity atom with mass $m_b$, and $V$ is the two-dimensional volume.
Here, the pairing order parameter is $\Delta_{\bf k}=\Delta k_{-}$, with $\Delta=[g_{\rm{ff}}/(2V)]\sum_{{\bf k}'}k'_{+}\langle a_{-{\bf k}'}a_{{\bf k}'}\rangle$, $k_{\pm}=k_{x} \pm ik_{y}$, and $g_{\rm{ff}}$ is the $p$-wave interaction between fermions in the background.
$g_{\rm{fi}}$ is the $s$-wave interaction rate between fermions and the impurity, which can be renormalized
following the standard procedure in two dimensions~\cite{Bruun2D2011,Parish2D2011-1,Parish2D2013,SOC2D2012,SOCpolaron2016}:
\begin{align}
\frac{1}{g_{\rm{fi}}} = -\frac{1}{V}\sum_{{\bf k}} \frac{1}{\epsilon_{{\bf k}} + \epsilon_{{\bf k}}^{b} + E_{b}}, \label{eq:renormalization2D}
\end{align} where $E_{b}$ is the two-body bound-state energy.
The natural units $\hbar = k_{B} = 1$ will be used throughout the paper.

The $p$-wave Fermi pairing superfluid at zero temperature can be described by the Bardeen-Cooper-Schrieffer (BCS)-type  wave function~\cite{Alicea2012}
\begin{align}
|\rm{BCS}\rangle_{p} &= \prod_{{\bf k}}\left(u_{{\bf k}} + v_{{\bf k}} a_{-{\bf k}}^{\dagger}a_{{\bf k}}^{\dagger} \right) |0\rangle, \\
u_{{\bf k}} &= e^{i\theta_{\bf k}}\sqrt{\frac{E_{\bf k} + (\epsilon_{\bf k} - \mu)}{2E_{\bf k}}}, \\
v_{{\bf k}} &= \sqrt{\frac{E_{\bf k} - (\epsilon_{\bf k} - \mu)}{2E_{\bf k}}},
\end{align}
where $\theta_{\bf k}=\text{arg}(k_{x} + ik_{y})$, $E_{{\bf k}} = \sqrt{(\epsilon_{\bf k} - \mu)^2 + |\Delta_{\bf k}|^2}$, and $|0\rangle$ is the vacuum state. The BCS-type wave function $|\rm{BCS}\rangle_{p}$ is the vacuum for Bogoliubov quasiparticles, with $\alpha_{{\bf k}}|\rm{BCS}\rangle_{p}=0$ and $\alpha_{{\bf k}} = u_{{\bf k}} a_{{\bf k}} + v_{{\bf k}} a_{-{\bf k}}^{\dagger}$.

\begin{figure}
\includegraphics[width=8cm]{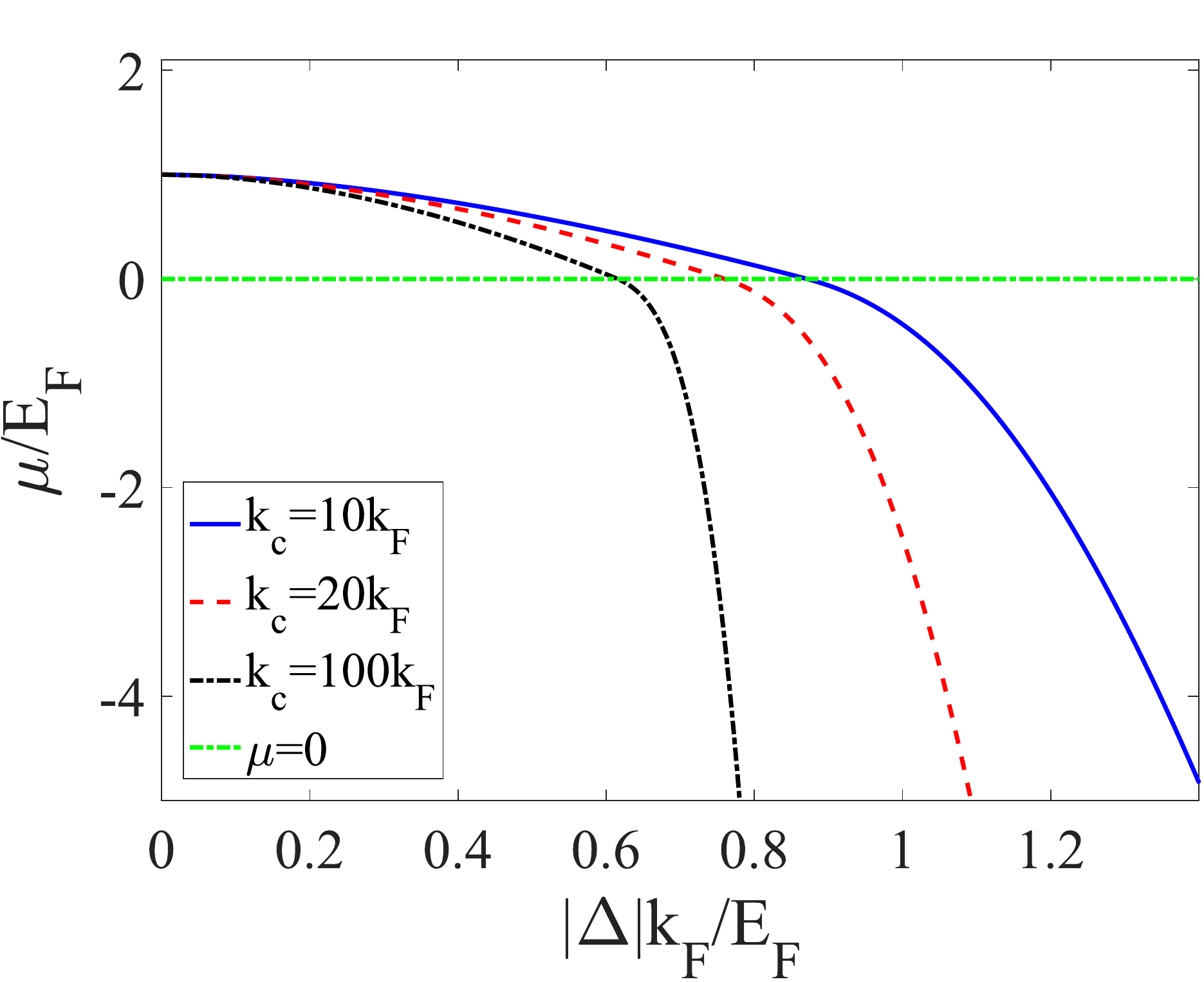}
\caption{(Color online) Chemical potential $\mu$ of the background $p$-wave superfluid as a function of the pairing order parameter $\Delta$. \label{fig:background}}
\end{figure}

The parameters characterizing the background $p$-wave superfluid are related through the gap and number equations,
\begin{align}
\frac{4}{g_{\rm{ff}}}&= -\frac{1}{V}\sum_{{\bf k}}\frac{k^{2}}{E_{{\bf k}}}, \\
n &= \frac{1}{V}\sum_{{\bf k}} |v_{{\bf k}}|^{2}, \label{eq:number}
\end{align} where the particle number density $n$ is given by $n=k_{F}^{2}/(4\pi)$ with the Fermi wave vector $k_{F}$. For the summations here, we adopt a high-momentum cutoff $k_c$, which is related to the short-range scattering parameter of the $p$-wave interaction. In Fig.~\ref{fig:background}, we show the chemical potential as a function of the pairing order parameter. For a given $k_c$, the chemical potential monotonically decreases with increasing $|\Delta|$, as the system changes from a topologically nontrivial phase with $\mu>0$ to a topologically trivial phase with $\mu<0$. In the following, we will use $\mu$ or $\Delta$ to characterize the background $p$-wave interaction strength.

\section{Polaron state}\label{3}

\begin{figure}
\includegraphics[width=8.5cm]{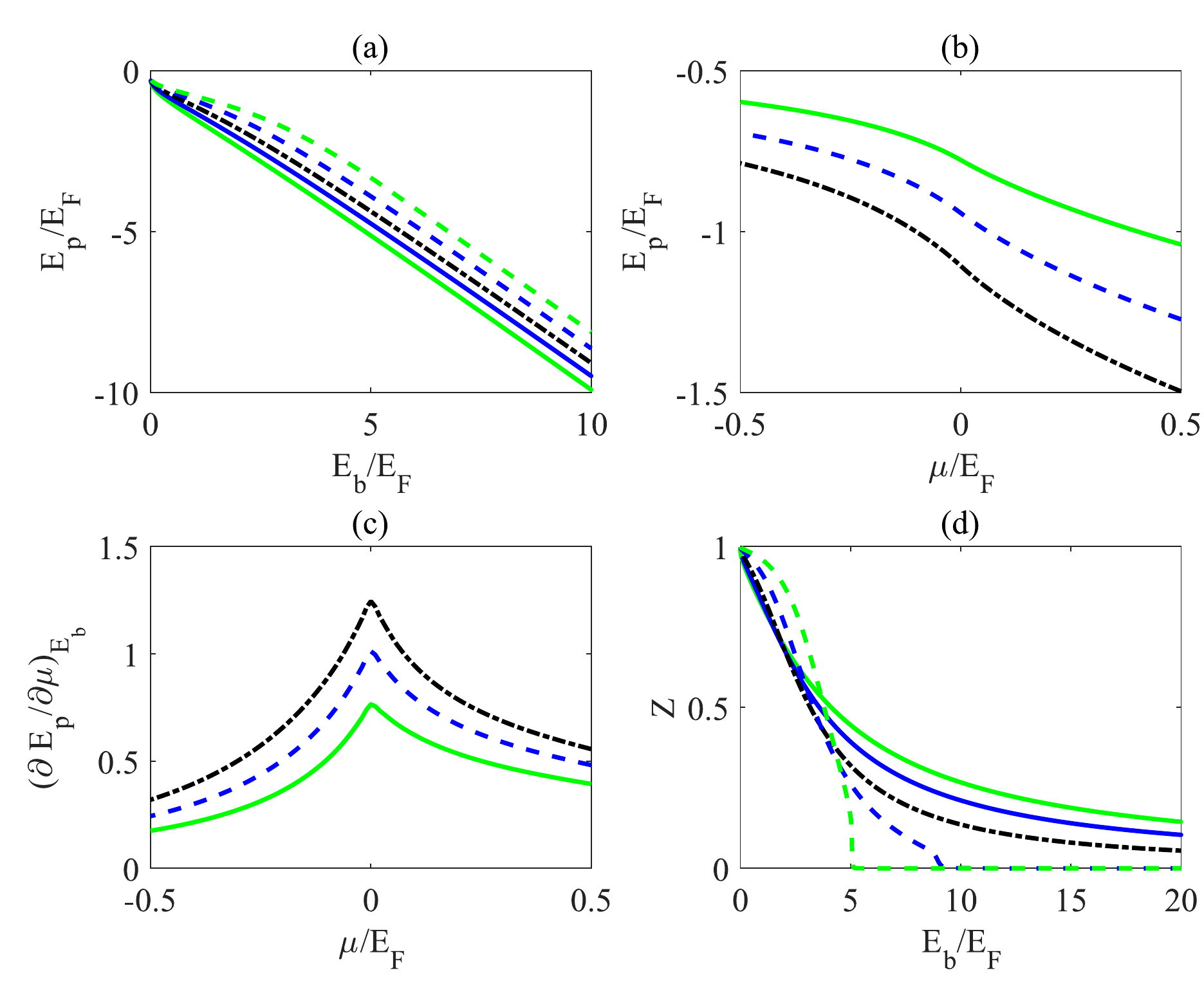}
\caption{(Color online) (a) The lowest branch of the polaron energy versus $E_{b}$. (b) The lowest branch of the polaron energy versus $\mu$. (c) The first derivative of the lowest branch of the polaron energy with respect to $\mu$. (d) Impurity residue. The parameters are $Q=0$, $m_b=m$, and momentum cutoff $k_c=20k_F$. In (a) and (d), the solid green line is for $\mu=0.5E_F$, the solid blue line is for $\mu=0.2E_F$, the dashed-dotted black line is for $\mu=0$, the dashed blue line is for $\mu=-0.2E_F$, and the dashed green line is for $\mu=-0.5E_F$. In (b) and (c), the solid green line is for $E_b=0.5E_F$, the dashed blue line is for $E_b=0.75E_F$, and the dashed-dotted black line is for $E_b=E_F$. \label{fig:Ep}}
\end{figure}

We adopt the Chevy-type ansatz to characterize the polaron state~\cite{Chevy2006,Yi2015}:
\begin{align}
&~~|P\rangle_{{\bf Q}}= \nonumber\\
& \left(\psi_{{\bf Q}}b^{\dagger}_{{\bf Q}} + \frac{1}{2}\sum_{{\bf k},{\bf k}'}\psi_{{\bf k},{\bf k}'}b^{\dagger}_{{\bf Q}-{\bf k}-{\bf k}'}\alpha^{\dagger}_{{\bf k}'}\alpha^{\dagger}_{{\bf k}} \right)|\rm{BCS}\rangle_{p}, \label{eq:PolaronAnsatz}
\end{align}
where $\psi_{{\bf Q}}$ and $\psi_{{\bf k},{\bf k}'}$ are polaron wave functions, $\psi_{{\bf k},{\bf k}'}=-\psi_{{\bf k}',{\bf k}}$ due to the $p$-wave symmetry, ${\bf Q}$ indicates the center-of-mass momentum of the polaron. The second term in the brackets, which effectively describes impurity-induced pair breaking in the superfluid and is therefore trimerlike, includes contributions from excitations like $a^{\dagger}_{{\bf k}'}a_{-{\bf k}}$, $a_{-{\bf k}'}a^{\dagger}_{{\bf k}}$, $a^{\dagger}_{{\bf k}'}a^{\dagger}_{{\bf k}}$, or $a_{-{\bf k}'}a_{-{\bf k}}$. Here, we only keep excitations to the lowest order. The existence of a trimerlike term in the polaron wave function is unique for a pairing-superfluid background.

The ground-state solution can be obtained by minimizing $E_{p}= {_{{\bf Q}}}\langle P|(H-\mu N)|P\rangle_{{\bf Q}} - E_{\rm{BCS}}$, where $E_p$ is the polaron energy and the BCS ground-state energy is
\begin{align}
E_{\rm{BCS}}= \frac{1}{2}\sum_{{\bf k}}[(\epsilon_{\bf k} - \mu) - E_{\bf k}] + \frac{|\Delta|^2}{4V}\sum_{{\bf k}}\frac{k^{2}}{E_{{\bf k}}}.
\end{align}

Following a similar derivation as in Ref.~\cite{Yi2015}, we have
\begin{align}
\left(\frac{V}{g_{\rm{fi}}} - \sum_{{\bf k}'}\frac{|u_{{\bf k}'}|^{2}}{A_{{\bf k},{\bf k}'}} \right)A_{\bf k}
&= \frac{v_{\bf k}\sum_{{\bf k}'}v_{{\bf k}'}^{*}A_{{\bf k}'}}{E_p-\epsilon_{\bf Q}^{b}} - u_{\bf k}\sum_{{\bf k}'}\frac{u_{{\bf k}'}^{*}A_{{\bf k}'}}{A_{{\bf k},{\bf k}'}}, \label{eq:ClosedEquationA}
\end{align}
where
\begin{align}
A_{{\bf k},{\bf k}'}& = E_p - E_{{\bf k}} - E_{{\bf k}'} - \epsilon_{{\bf Q}-{\bf k}-{\bf k}'}^{b}, \label{Akk_new}\\
A_{\bf k}&=g_{\rm{fi}}\left(v_{\bf k} \psi_{\bf Q} + \sum_{{\bf k}''}u^{*}_{{\bf k}''}\psi_{{\bf k},{\bf k}''}\right). \label{Ak_new}
\end{align}

The coefficients in the polaron ansatz (\ref{eq:PolaronAnsatz}) give the impurity residue~\cite{Yi2015},
\begin{align}
Z = \frac{|\psi_{\bf Q}|^{2}}{|\psi_{\bf Q}|^{2} + \frac{1}{4}\sum_{\bf k,\bf k'} |\psi_{\bf k,\bf k'}|^{2}}. \label{eq:PolaronResidue}
\end{align}
The impurity residue signals the weight of bare impurity in the polaron excitation. From the ansatz wave function Eq.~(\ref{eq:PolaronAnsatz}), it is clear that the polaron would become trimer-like when $Z$ approaches zero.

We numerically solve Eq.~(\ref{eq:ClosedEquationA}) and get the lowest-energy polaron branch as shown in Figs.~\ref{fig:Ep}(a) and \ref{fig:Ep}(b). The polaron binding energy $|E_p|$ increases with increasing impurity-fermion interaction strength or with increasing fermion-fermion interaction.
We then plot the first derivative of the $E_p$ with respect to $\mu$. As shown in Fig.~\ref{fig:Ep}(c), kinks appear at $\mu=0$ regardless of the impurity-fermion interaction, which indicates a discontinuity in the second derivative of the polaron energy. This is consistent with the fact that the topological phase transition in the background is a third-order phase transition. Thus the information of the background topological phase transition is carried over to the polaron excitation, which can serve as a probe for the phase transition.

In Fig.~\ref{fig:Ep}(d), we show the evolution of the impurity residue as a function of $E_b$. When the Fermi superfluid is in the topologically trivial strongly interacting regime ($\mu<0$), the impurity residue drops precipitously over a small range of $E_b$, such that the residue $Z$ essentially vanishes at large impurity-fermion interactions. This is a clear signature of polaron to trimer crossover, first studied in the impurity problem for a background of $s$-wave Fermi superfluid~\cite{Yi2015}. However, the polaron to trimer crossover becomes much smoother in the topologically nontrivial weak-interacting regime ($\mu>0$), as the residue monotonically decreases but remains finite even at large $E_b$.

\begin{figure}
\includegraphics[width=8.8cm]{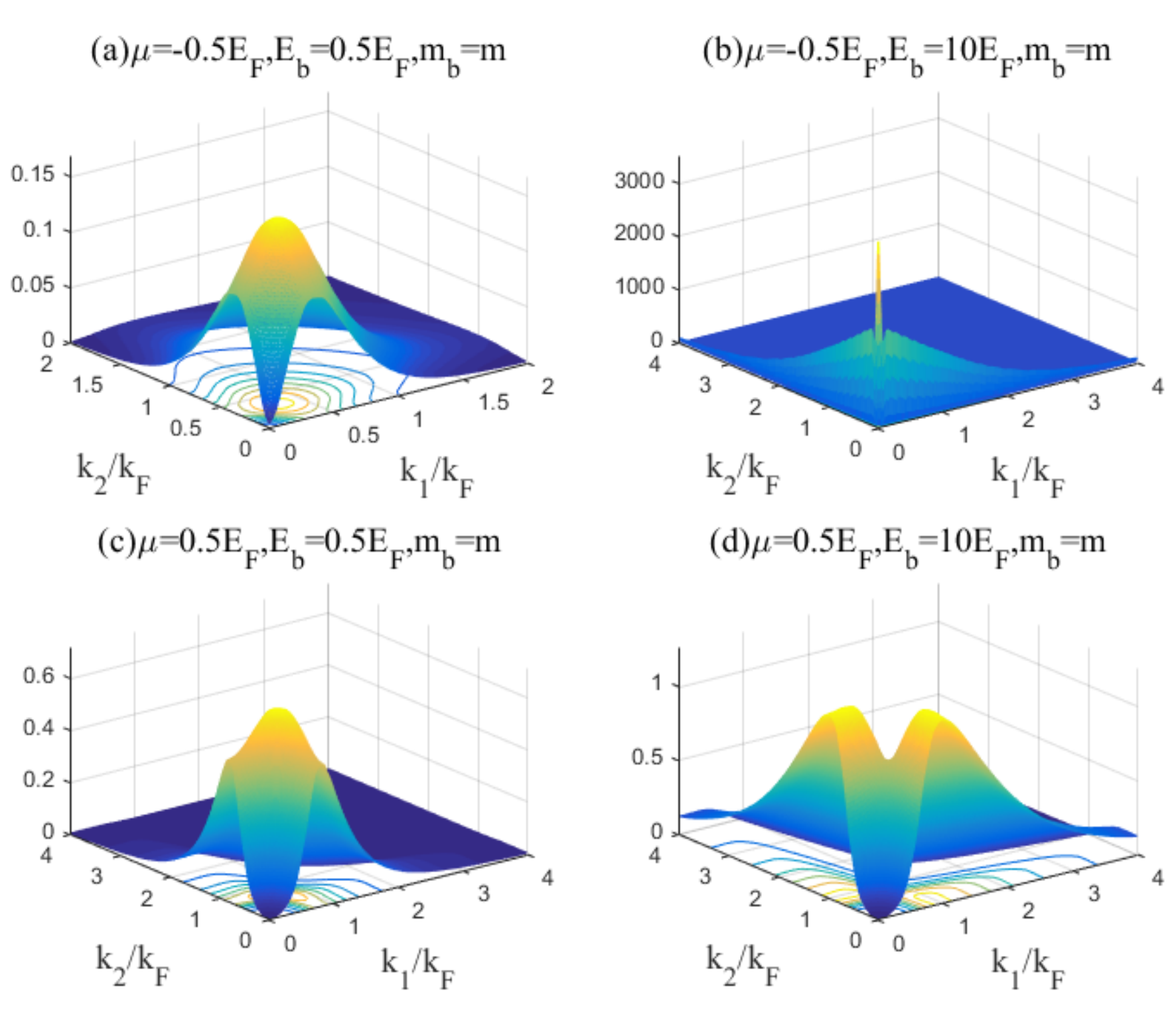}
\caption{(Color online) Momentum distribution of the trimer probability $|\psi_{{\bf k}_1,{\bf k}_2}|^2$ in the $(k_1,k_2)$ plane, where the polar angles $\theta_1$ and $\theta_2$ of ${\bf k}_1$ and ${\bf k}_2$ have been integrated for (a) $E_b=0.5E_F$ with $\mu=-0.5E_F$, (b) $E_b=10E_F$ with $\mu=-0.5E_F$, (c) $E_b=0.5E_F$ with $\mu=0.5E_F$, and (d) $E_b=10E_F$ with $\mu=0.5E_F$. The other parameters are $Q=0$, $m_b=m$, and momentum cutoff $k_c=20k_F$.} \label{fig:MomentumDistribution_ksum}
\end{figure}

The polaron-trimer crossover is more apparent in the momentum-space probability distribution of the wave function $\psi_{{\bf k}_1,{\bf k}_2}$.
In Figs.~\ref{fig:MomentumDistribution_ksum}(a) and \ref{fig:MomentumDistribution_ksum}(b), we show the angular-integrated momentum-space distribution $\int d\theta_1d\theta_2|\psi_{{\bf k}_1,{\bf k}_2}|^2$ for $\mu<0$, where $\theta_1$ and $\theta_2$ are the polar angles of ${\bf k}_1$ and ${\bf k}_2$, respectively. When the impurity-fermion interaction is small, the distribution is localized in momentum space [see Fig.~\ref{fig:MomentumDistribution_ksum}(a)]. This suggests an extended polaronlike spatial wave function. When the impurity-fermion interaction is large, the distribution becomes extended in momentum space, as shown in Fig.~\ref{fig:MomentumDistribution_ksum}(b). This suggests a spatially localized trimerlike wave function. In contrast, for $\mu>0$, the momentum distribution of the wave function is always localized, regardless of the impurity-fermion interaction as shown in Figs.~\ref{fig:MomentumDistribution_ksum}(c) and \ref{fig:MomentumDistribution_ksum}(d). Further, we have numerically checked that the probability $\int k_{1}k_{2}dk_{1}dk_{2}|\psi_{{\bf k}_1,{\bf k}_2}|^2$ always peaks at $\theta_1=\theta_2+\pi$.

\section{Molecular state}\label{4}

\begin{figure}
\includegraphics[width=8.5cm]{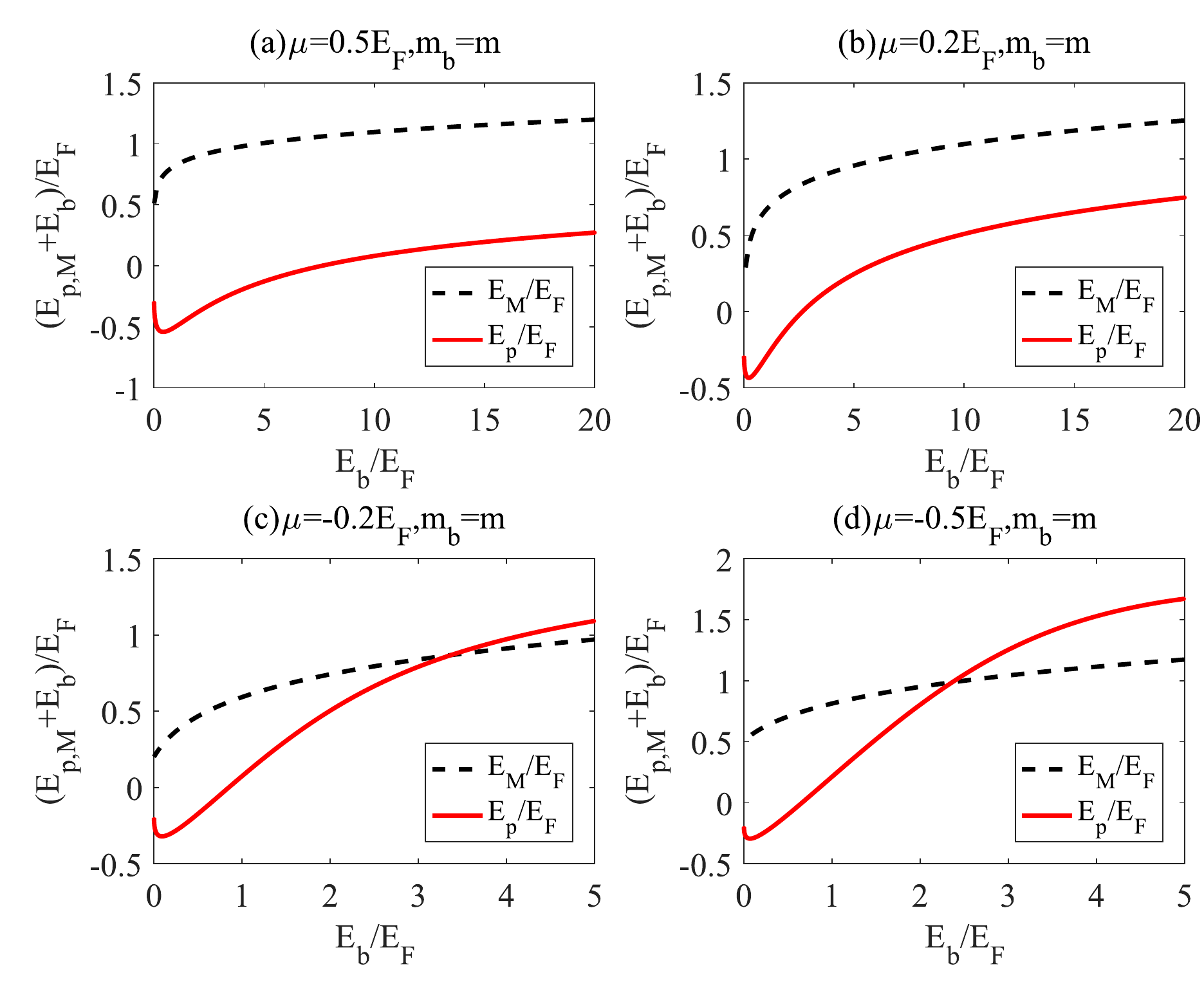}
\caption{(Color online) Molecular and polaron energies in the $Q=0$ sector with (a) $\mu=0.5E_F$; (b) $\mu=0.2E_F$; (c) $\mu=-0.2E_F$; (d) $\mu=-0.5E_F$. Here, we choose $m_b=m$ and $k_c=20k_F$. The red solid curve is the lowest branch of the polaron energy and the black dashed curve is the molecular energy. \label{fig:PM}}
\end{figure}

For the impurity problem in a non-interacting Fermi gas, a polaron to molecule transition occurs when the impurity-fermion interaction increases. In two dimensions, such a transition can be accurately captured by considering polaron and molecular wave functions each dressed by a single pair of particle-hole excitation~\cite{Bruun2D2011,Parish2D2011-1}. To determine the ground state of the impurity, we should also compare the energy of a molecular state with that of the polaron state.

We write the molecular ansatz wave function as~\cite{Parish2D2013,Yi2015}
\begin{align}
|M\rangle_{{\bf Q}} = \sum_{{\bf k}}\phi_{{\bf k}}^{(M)}b^{\dagger}_{{\bf Q}-{\bf k}}\alpha^{\dagger}_{{\bf k}} |\rm{BCS}\rangle_{p},\label{eq:mol}
\end{align}
where ${\bf Q}$ is the center-of-mass momentum of the molecule. Physically, we are considering the binding of the impurity and a Bogoliubov quasiparticle into a tightly confined dimer. In the weak-coupling limit $\Delta\rightarrow 0$, the wave function above reduces to that of a bare molecular state.

The ground-state solution can be obtained by minimizing $E_{M}= {_{{\bf Q}}}\langle M|(H-\mu N)|M\rangle_{{\bf Q}} - E_{BCS}$, where $E_M$ is the molecular energy. The closed equation reads
\begin{align}
\frac{1}{g_{\rm{fi}}}
= \frac{1}{V} \sum_{{\bf k}} \frac{|u_{{\bf k}}|^{2}}{E_{M} - E_{{\bf k}} - \epsilon_{{\bf Q}-{\bf k}}^{b}},
\end{align}
where $g_{\rm{fi}}$ needs to be renormalized following Eq.~(\ref{eq:renormalization2D}).

In Fig.~\ref{fig:PM}, we show the lowest branch of the polaron energy as well as the molecular energy. In the topologically trivial regime, a polaron to molecule transition occurs. By comparing with Fig.~\ref{fig:Ep}(d), we find that the polaron to molecule transition happens prior to the polaron-trimer crossover region. Therefore, the trimer-like state is metastable. In contrast, when the Fermi superfluid is topological, we find no polaron to molecule transitions. However, similar to the case of two-dimensional Fermi polaron, where the polaron to molecule transition only appears when one considers dressed molecules with higher-order particle-hole fluctuations~\cite{Bruun2D2011,Parish2D2011-1}, we expect that a polaron to molecule transition should exist against a background of topological superfluid when higher-order fluctuations in the molecular sector are taken into account. Specifically, one needs to go beyond Eq.~(\ref{eq:mol}) and consider more pairs of quasiparticle excitations. Nevertheless, the fact that we find polaron to molecule transitions in the strongly interacting regime by considering the bare molecular state alone suggests that pairing interaction in the background actually destabilizes the polaron state and effectively facilitates the binding of impurity and fermion.

\section{Summary}\label{5}

By using the Chevy-type variational wave functions, we study the impurity problem of a $p+ip$ Fermi topological superfluid across the topological phase transition. The topological phase transition gives rise to kinks of the polaron-enegy derivatives, which can serve as an external probe for the phase transition. We then discuss the interplay of fermion-fermion and impurity-fermion interactions on the polaron-trimer crossover as well as the polaron-molecule transition in the system. Our results reveal interesting competitions between various many-body states in an interacting topological system.

\section*{Acknowledgements}
We thank Ming Gong, Lijun Yang, Tian-Shu Deng, and Jing-Bo Wang for useful discussions. W.Y. acknowledges Jiali Lu for his early contributions.
X.C. acknowledges support from the National Key Research and Development Program of China (Grants No. 2018YFA0307600 and No. 2016YFA0300603), and the National Natural Science Foundation of China (Grants No. 11622436, No. 11421092, and No. 11534014).
W.Y. acknowledges support from the National Key Research and Development Program of China (Grants No. 2016YFA0301700 and No. 2017YFA0304100), and the National Natural Science Foundation of China (Grant No. 11522545).
F.Q. acknowledges support from the National Key Research and Development Program of China (Grant No. 2017YFA0304800), the National Natural Science Foundation of China (Grant No. 11404106), and the project funded by the China Postdoctoral Science Foundation (Grant No. 2016M602011).

\end{document}